\documentstyle[fleqn,11pt]{article}
\date{}

\begin{document}

\title{\Large \bf The Pion Mass Formula}
\author{{Reginald T. Cahill  and      Susan M. Gunner
  \thanks{E-mail: Reg.Cahill@flinders.edu.au,
 Susan.Gunner@flinders.edu.au}}\\
{Department of Physics, Flinders University}\\ 
{ GPO Box 2100, Adelaide 5001, Australia}\\{hep-ph/9602240}}

\maketitle

\begin{center}
\begin{minipage}{120mm}
\vskip 0.6in
\begin{center}{\bf Abstract}\end{center}
{The often used Gell-Mann-Oakes-Renner (GMOR) mass formula for Nambu-Goldstone   bosons
in QCD, such as the pions, involves the condensate $<\overline{q}q>$, $f_{\pi}$ and the quark
current masses. Within the context of the Global Colour Model (GCM) for QCD a manifestly
different  formula was recently found  by Cahill and Gunner.  Remarkably Langfeld and
Kettner have shown the two formulae to be equivalent. Here we explain that the above recent
analyses refer to the GCM constituent pion and not the exact GCM pion. Further, we suggest
that the GMOR formula is generic. We generalise the Langfeld-Kettner identity to
include the full response of the constituent quark propagators to the presence of a  
non-zero (and running) quark current mass.


}
\end{minipage} \end{center}

\newpage

\vskip 1.5cm
\par
     
\noindent {\bf 1. Introduction}

The properties of the pion continue to be the subject of considerable theoretical and
experimental interest in QCD studies.  The pion is an (almost) massless Nambu-Goldstone (NG)
boson and its properties are directly associated with dynamical chiral symmetry breaking and
the underlying quark-gluon dynamics. Recently there has been renewed interest in the mass
formula for the pion  \cite{CG95b,FR96,Lang} and the relationship with the well known
Gell-Mann-Oakes-Renner (GMOR) \cite{GMOR} mass formula, as in   (1) and (2).  Here we  extend
the study of  these relationships and show how one must carefully appreciate the different
quantum field theoretic approaches that are actually being employed, often without explicit
exposition.  

One expects that
there should be some perturbative expression for the almost NG boson   pion mass in terms
of the small quark current masses which   is built upon the underlying non-perturbative
chiral-limit quark-gluon dynamics. While the relation of the low pion mass to the breaking
of chiral symmetry dates back to the current algebra era and PCAC
\cite{GMOR}, the often used implementation in QCD has the form,  
\begin{equation}
m_{\pi}^2= \frac{(m_u+m_d)\rho}{f_{\pi}^2}
\label{eq:1}\end{equation}
where the integral $\rho=<\overline{q}q>$ is the so-called condensate parameter. For $N_c=3$ 
\begin{equation}
\rho=N_ctr(G(x=0))=12\int\frac{d^4q}{(2\pi)^4}\sigma_s(q^2),
\label{eq:2}\end{equation}
and $f_{\pi}$ is the usual pion decay constant. (Note: our definition for $\rho$ has an
unconventional sign). 
In (2)  $\sigma_s(s) \equiv \sigma_s(s;0)$   is the chiral limit  scalar part of the  quark 
propagator which, utilizing only the Lorentz
structure (and for later reference $\sigma_v(s) \equiv
\sigma_v(s;0)$), we can write in full generality as   \begin{equation}  
G(q;m)=(iA(s;m)q.\gamma+B(s;m)+m(s))^{-1}=-iq.\gamma\sigma_v(s;m)+\sigma_s(s;m), 
\label{eq:3}\end{equation}
from which we  easily deduce that
\begin{equation}
\sigma_s(s;m)=\frac{B(s;m)+m(s)}{sA(s;m)^2+(B(s;m)+m(s))^2},
\label{eq:3a}\end{equation}
\begin{equation}
\sigma_v(s;m)=\frac{A(s;m)}{sA(s;m)^2+(B(s;m)+m(s))^2}.
\label{eq:3b}\end{equation}
Here $m(s)$ is the running quark current mass, but  only the combination
$B(s;m)+m(s)$ appears.  
We note  that the expression for $\rho$ in (2) is
divergent  in QCD, because for  $s \rightarrow \infty$ 
$B(s)$ decreases like  $1/(s\mbox{ln}[s/\Lambda^2]^{1-\lambda})$
where $\lambda = 12/(33-2N_f)$ and $\Lambda$ is the QCD scale parameter.   Some
integration cutoff  is usually
 introduced, and    the  values of $m$ and
$<\overline{q}q>$ are  quoted as being relative to this cutoff momentum, often $1
GeV$. The GMOR relation has been considered in various  approaches, such as  operator
product expansions (OPE)
\cite{OPE},  QCD sum rules \cite{Rein85,Nar89} and recently, finite energy sum rules and
Laplace sum rules
\cite{Bijnens}.

In \cite{CG95b} a new mass formula for the pion mass was derived. The analysis in
\cite{CG95b}  exploited the  intricate interplay between the constituent pion Bethe-Salpeter
equation (BSE) and the non-linear Dyson-Schwinger equation (DSE) for the constituent quarks,
resulting in the new expression
\begin{equation} m_{\pi}^2=\frac{24m}{f_{\pi}^2}  \int
\frac{d^4q}{(2\pi)^4}\left(\epsilon_s(s) \sigma_s(s)+s\epsilon_v(s)
\sigma_v(s)\right)c(s)+O(m^2),  
\label{eq:20a}\end{equation} 
where $c(s)$ is a naturally arising cutoff function 
\begin{equation} c(s)=\frac{B(s;0)^2}{sA(s;0)^2+B(s;0)^2}.
\label{eq:21a}\end{equation} 
Here $\epsilon_s(s)$ and $\epsilon_v(s)$ are functions which specify the response of the
constituent quark propagator to the turning on of the quark current mass; see (\ref{eq:12})
and (\ref{eq:13}).  Note that the GMOR mass formula (\ref{eq:1}) and (\ref{eq:2}) appear to
be manifestly different from the new expression in  (\ref{eq:20a}). 
However Langfeld and Kettner \cite{Lang}  have shown, by further analysis of the DSE for the
constituent quark propagator, and ignoring for  simplicity the $\epsilon_v(s)$
vector response term, that the two mass formulae are  equivalent, even though the
integrands are indeed different. 

Here we  first demonstrate that in the quantum field theoretic analyses  different concepts
are often being used and confused in the literature. In this respect we carefully distinguish
between the  constituent pion and the full or exact pion, and their relevant mass
expressions.  Little detailed progress has been made in the exact analysis of QCD,
and so we use the Global Colour Model (GCM) to illustrate these differences. Further we
extend the Langfeld-Kettner identity \cite{Lang} to   include 
the vector response function  $\epsilon_v(s)$ and a quark runnning current mass function
$m(s)$. We show that the new mass formula can indeed be written in the GMOR
form, with both  (\ref{eq:20a}) and (\ref{eq:1}) each now generalised to include a running
current mass.  

To be clear  we note that this report contains no analysis of the mass formula for
the full pion in QCD, or even in the GCM. However if the GMOR relation is also the correct
QCD result up to
$O(m)$, then that would indicate that the GMOR relation is in fact a {\em generic} form that
arises whether we are dealing with the full pion or with the constiutent pion, and whether
we are analysing QCD or some approximation scheme to QCD, such as the GCM, provided we
carefully preserve the dynamical  consequences  of the dynamical breaking of chiral symmetry
and its activation  by the underlying quark-gluon dynamics.  We also note that the GMOR
formlua has been `derived' in the past, but such analyses in general appear to have brushed
over  the various subtleties presented herein. Ref.\cite{CG95b}, in its appendix, 
illustrated this by using one example of an incorrect derivation leading to the GMOR
relation.

 We exploit the GCM  \cite{RTC} of QCD which has proven to be remarkably successful in
modelling low energy QCD, as discussed  in sec.2.  It should be
noted that the GCM generates constituent hadrons which necessarily have the form of 
ladder states. The non-ladder  diagrams  then arise from hadronic functional
integrations over constituent ladder hadrons.  In sec.3 we show the difference between the
full or exact and the constiutent pion. This involves the use of effective actions and the
fact that these effective actions  refer to constituent hadrons. In sec.4 the effective
action for the chiral limit constituent pion is discussed.  In sec.5   the constituent quark
propagators are given. They arise as the Euler Lagrange  equations  of the hadronised
effective action for the GCM. They define the constituent quarks.  Fluctuations about the
minimum action  configuration introduce constituent mesons, and these are described by 
ladder BSEs. {\em Ad hoc} alterations to these equations can introduce double counting
problems.   The full (observable) states are produced by dressing each of the constituent
states by other states, as is made clear by the functional integral formalism in sec.2. In
sec.6 the constituent pion mass formula, (\ref{eq:20a}), is derived, but here generalised
to include  the quark running mass.  In sec.7 the Langfeld-Kettner identity is generalised
to include the vector reponse function and a quark running current mass. This identity
leads from the new formula in (\ref{eq:20a}) to the  GMOR formula in (\ref{eq:1}).

\vspace{10mm}
\noindent {\bf 2. The Global Colour Model of QCD}

An overview and an insight into the nature of the non-perturbative low energy hadronic regime
of  QCD is provided by the functional integral hadronisation of QCD \cite{RTC}.
In the functional integral approach correlation functions of QCD are defined by 
\begin{equation} {\cal G}(..,x,...)=\int{\cal D}\overline{q}{\cal D}q{\cal
DA}....q(x).....exp(-S_{QCD}[A,\overline{q},q])
\label{eq:2.1}\end{equation}
  the kernel of  which includes (not shown)  gluon string structures that render ${\cal
G}(..,x,...)$ gauge invariant.  One example of  (\ref{eq:2.1}) would be the pion correlation
function, which may be defined by the connected part of

\begin{equation} {\cal G}_{\pi}(x,y,z,w)=\int{\cal D}\overline{q}{\cal D}q{\cal
D}A....\overline{q}(x)i\gamma_5\tau_iq(y)..\overline{q}(z)i\gamma_5\tau_iq(w)..
exp(-S_{QCD}[A,\overline{q},q]).
\label{eq:2.2}\end{equation}
 The $\pi\pi$ scattering amplitude, for example,  is also defined by such a functional
integral.  The pion mass is defined by the position of the pole, wrt the centre-of-mass (cm)
momentum, of the Fourier transform of the translation invariant amplitude ${\cal G}_{\pi}$.
Apart from lattice computations a direct computation of these functional integrals  is not
attempted. Amplitudes, such as (\ref{eq:2.2}), when the  on-mass-shell conditions are
imposed, define the observables of QCD, such as the pion.  Theoretical analysis of these
amplitudes proceeds by more circumspect techniques, some of which we clarify here.

The correlation functions, as  in (\ref{eq:2.2}), may be extracted from the generating
functional of QCD, $Z_{QCD}[\overline{\eta},\eta,..]$, defined in (\ref{eq:2.2b}).
However the interactions of low energy hadronic physics, such as $\pi\pi$ scattering, are
 known  to be well described by effective actions which refer only to
hadronic states, although the various parameters in these effective actions could only be
determined by fitting to experimental data.  Hence  we  expect  that the functional
integrals, such as (\ref{eq:2.2}),  should also be extractable from a hadronic
functional integral, as is indeed possible 
\begin{equation}
Z_{QCD}[\overline{\eta},\eta,..]=\int {\cal D}\overline{q}{\cal D}q{\cal
D}Aexp(-S_{QCD}[A,\overline{q},q]+\overline{\eta}q+
\overline{q}\eta)
 \label{eq:2.2b}\end{equation}
 \begin{equation}\mbox{\ \ \ \ \ \ \ \ \ \ \ \ \ \ \ \  } 
\approx \int{\cal D}\pi{\cal D}\overline{N}{\cal
D}N...exp(-S_{had}[\pi,..,\overline{N},N,..]+J_{\pi}[\overline{\eta},\eta]\pi+..)
\end{equation}
\begin{equation} \mbox{\ \ \ \ \ \ \ \ \ \ \ \ \ \ \ \ } 
= Z_{had}[J_{\pi}[\overline{\eta},\eta],...],
\label{eq:2.3}
\end{equation}
which produces a hadronic generating
functional, $Z_{had}[J_{\pi}[\overline{\eta},\eta],...]$, in which  source terms for the
various  hadrons  are naturally induced.  A partial derivation \cite{RTC} of this functional
transformation proceeds as follows. First, and not showing source terms for convenience, 
the gluon integrations are formally performed (ghosts also not shown)
\begin{eqnarray*}\int {\cal D}\overline{q}{\cal D}q{\cal
D}Aexp(-S_{QCD}[A,\overline{q},q])\end{eqnarray*}
\begin{eqnarray*} \mbox{\ \ \ \ \ \ \ \ \ \ \ \ }= \int {\cal D}\overline{q}{\cal D}q
exp(-\int 
\overline{q}(\gamma . \partial+{\cal M})q +
\end{eqnarray*}
\begin{equation}\mbox{\ \ \ \ \ \ \ \ \ \ \ \ \ \ \ \ \ \ \ \ \ \ \ }
+\frac{1}{2}\int
j^a_{\mu}(x)j^a_{\nu}(y)D_{\mu\nu}(x-y)+\frac{1}{3!}\int
j^a_{\mu}j^b_{\nu}j^c_{\rho}D^{abc}_{\mu\nu\rho}+..)\label{eq:2.4}\end{equation}where
$j^a_{\mu}(x)=\overline{q}(x)\frac{\lambda^a}{2}\gamma_{\mu}q(x)$, and $D_{\mu\nu}(x)$ is
the exact pure gluon propagator 
\begin{equation}D_{\mu\nu}(x-y)=\int {\cal D}AA^a_{\mu}(x)A^a_{\nu}(y)exp(-S_{QCD}[A,0,0]).
\label{eq:2.5}\end{equation} A variety of techniques for computing $D_{\mu\nu}(x)$ exist,
such as the gluonic  DSE  \cite{Pennington} and lattice simulations
\cite{Marenzoni93}. The terms of higher order than  quartic in the quark
fields   are beyond our ability to retain in the analysis.   Dropping these terms, beginning
with 
$D^{abc}_{\mu\nu\rho},..$,  defines the GCM. Apart from lattice modellings of QCD no
theoretical analyses   have incorporated these  higher order correlation functions in  hadron
studies. The remarkable success of the GCM suggests, for unknown reasons, that these terms
are ineffective, at least in (colour singlet) hadronic states. Of course
$D^{abc}_{\mu\nu\rho},..$  play an important role in the DSE approach \cite{Pennington} to
estimating  (\ref{eq:2.5}). This GCM truncation is equivalent to using a quark-gluon field
theory with the action
\begin{equation} S_{GCM}[\overline{q},q,A^a_{\mu}]=\int \left( 
\overline{q}(\gamma . \partial+{\cal M}(i\partial)
+iA^a_{\mu}\frac{\lambda^a}{2}\gamma_{\mu})q
 +\frac{1}{2} A^a_{\mu}D^{-1}_{\mu\nu}(i\partial)A^a_{\nu} \right).\label{eq:2.6}
\end{equation} 
Here $D_{\mu\nu}^{-1}(p)$ is the matrix inverse of $D_{\mu\nu}(p)$, which in turn is
the Fourier transform of $ D_{\mu\nu}(x)$. This action has only a global colour symmetry,
unlike the local colour symmetry that characterises QCD. The gluon self-interactions, which
arise as a consequence of  the local colour symmetry in (\ref{eq:2.5}),  lead to 
$D_{\mu\nu}^{-1}(p)$ being  non-quadratic.  Its precise form is  unknown, but see
\cite{Pennington} and \cite{Marenzoni93}.   In the GCM a
general non-quadratic form for 
$D_{\mu\nu}$ is  retained,  modelling this significant  property of QCD. One can even
attempt to extract $D_{\mu\nu}(p)$ from meson data \cite{CG95a}.  The quark current masses
${\cal M}(s)=\{m_u(s), m_d(s),..\}$ in (\ref{eq:2.6})  are allowed to run.  In momentum space
this leads to the $s$-dependence of $m(s)$ in (\ref{eq:7}).

Hadronisation
\cite{RTC}  involves a sequence of functional integral  calculus changes of
variables involving, in part, the transformation to bilocal meson and diquark fields, and
then to the usual local meson and baryon fields (sources not shown). 

\begin{eqnarray*}  Z=\int {\cal D}\overline{q}{\cal D}q{\cal
D}Aexp(-S_{QCD}[A,\overline{q},q]+\overline{\eta}q+
\overline{q}\eta)
\end{eqnarray*}
\begin{eqnarray*}\mbox{\ \ }\approx\int {\cal D}\overline{q}{\cal D}q{\cal
D}Aexp(-S_{GCM}[A,\overline{q},q]+\overline{\eta}q+
\overline{q}\eta)   \mbox{\ \ \ \ (GCM truncation)}
\end{eqnarray*}  
\begin{equation}
\mbox{\ \ } =\int D{\cal B}D{\cal D}D{\cal
D}^{\star}exp(-S_{bl}[{\cal B}, {\cal D}, {\cal D}^{\star}]) \mbox{\ \ \ \ (bilocal fields)}
\label{eq:2.7}\end{equation} 
\begin{equation} \mbox{\ \ } =\int{\cal D}\pi...{\cal D}\overline{N}{\cal
   D}N...exp(-S_{had}[\pi,...,\overline{N},N,..]) 
  \mbox{\ \ \ \ (local fields) }.\label{eq:2.8}\end{equation} 
The derived hadronic  action that finally emerges from this action sequencing, to low order
in fields and derivatives, has the form
\begin{eqnarray*} \lefteqn{ S_{had}[\pi,...,\overline{N},N,..] = }\\&
 &\mbox{\ \ \ \ \ \ \ \ \ \ \ \ \ \ }\int d^4x tr\{\overline{N}(\gamma.\partial+m_N+  \Delta
m_N-  m_N\surd 2i\gamma_5\pi^aT^a+..)N\}+\\ & &\mbox{\ \ \ \ \ \ \ \ \ \ }+\int
d^4x\left[ \frac{f_{\pi}^2}{2} [(\partial_{\mu}\pi)^2+m_{\pi}^2\pi^2] +
\frac{f_{\rho}^2}{2}[ -
\rho_{\mu}\Box\rho_{\mu} +(\partial_{\mu}\rho_{\mu})^2 +m_{\rho}^2\rho_{\mu}^2]+\right.\\   
& &\left.\mbox{\ \ \ \ \ \ \ \ \ \ }+\frac{f_{\omega}^2}{2}[\rho
\rightarrow \omega]-f_{\rho}f_{\pi}^2g_ {\rho\pi\pi}\rho_{\mu}.\pi \times\partial _{\mu}\pi
-if_{\omega}f_{\pi}^3\epsilon_{\mu\nu\sigma\tau}\omega_{\mu}\partial_{\nu}
 \pi . \partial_{\sigma}\pi\times \partial_{\tau}\pi+\right.\\ & &\left.\mbox{\ \ \ \ \ \ \
\ \ \ }-if_{\omega}f_{\rho}f_{\pi}
G_{\omega\rho\pi}\epsilon_{\mu\nu\sigma\tau}\omega_{\mu}\partial_{\nu}
\rho_{\sigma}.\partial_{\tau}\pi+\right.\end{eqnarray*}
\begin{equation}\left.\mbox{\ \ \ \ \ \ \ \ \ \ \ \ \ \ \ \ }+
\frac{\lambda i}{80\pi^2}\epsilon_{\mu\nu\sigma\tau}tr(\pi.F\partial_{\mu}
\pi.F\partial_{\nu}\pi.F\partial_{\sigma}\pi.F\partial_{\tau}\pi. F)
+......\label{eq:2.9}\right]\end{equation}

The bilocal fields in (\ref{eq:2.7}) arise naturally  and correspond to the fact that, for
instance, mesons are extended states. In (\ref{eq:2.2}) we can see that the pion
arises as a correlation function for two bilinear quark structures.  This
bosonisation/hadronisation arises   by functional integral calculus changes of variables
that are induced by generalised Fierz transformations that emerge from the colour, spin and
flavour structure of QCD.

 The final functional integration in (\ref{eq:2.8}) over the hadrons  give the
hadronic observables, and amounts to dressing each hadron by, mainly, lighter mesons.   
The basic insight is that the quark-gluon dynamics, in (\ref{eq:2.2}),  is
fluctuation dominated, whereas the hadronic functional integrations in  (\ref{eq:2.8}) are
not, and for example the meson dressing of bare hadrons is known to be almost perturbative.
In performing the change of variables essentially normal mode  techniques are used
\cite{RTC}. In practice this requires detailed numerical computation of the gluon
propagator, quark propagators, and  meson and baryon propagators. The mass-shell states of
the latter are determined by covariant Bethe-Salpeter  and  Faddeev equations. The Faddeev
computations are made feasible by using the diquark correlation propagators; the
diquarks being  quark-quark correlations within baryons. 

\vspace{15mm}
\noindent {\bf 3. Constituent Hadrons}

We now come to one of the main points.   Using the functional hadronisation 
we can write  ${\cal G}_{\pi}$ in the form (\ref{eq:2.2}) or, from (\ref{eq:2.8}), in the
form:

\begin{equation}
{\cal G}_{\pi}(X,Y)=\int{\cal D}\pi..{\cal D}\overline{N}{\cal
   D}N...\pi(X)\pi(Y)exp(-S_{had}[\pi,...,\overline{N},N,..]), 
\label{eq:2.10}\end{equation}
in which $X=\frac{x+y}{2}$ and $Y=\frac{z+w}{2}$ are cm  coordinates for the pion.  We note
that now the pion field appears in $S_{had}[\pi,...,\overline{N},N,..]$ which is in the
exponent of  (\ref{eq:2.10}), and it appears with an effective-action    mass parameter
$m_{\pi}$.  As we now discuss, it is important to clearly distinguish between this mass (and
the equations which define its value) and the pion mass that would emerge from
(\ref{eq:2.2}) or 
 (\ref{eq:2.10}).    Eqns.(\ref{eq:2.2}) or (\ref{eq:2.10})  define the observable pion
mass. Whereas the mass in (\ref{eq:2.9})  defines the constituent pion mass.   There is no
reason for these to be equal in magnitude, though they may well both be given by the
generic GMOR form.

  How do the constituent hadrons arise in (\ref{eq:2.8})?  In going from (\ref{eq:2.7})
to  (\ref{eq:2.8})  an expansion about the minimum of $S_{bl}[{\cal B}, {\cal D}, {\cal
D}^{\star}]$ is performed. First the minimum is defined by 
Euler-Lagrange equations (ELE)
\begin{equation}\frac{\delta S_{bl}}{\delta {\cal B}}=0
\mbox{ \ \ \ \ \ and  \ \ \ \ \ }
\frac{\delta S_{bl}}{\delta {\cal D}}=0\label{eq:2.11}.\end{equation} 
 These equations have solutions ${\cal B}\neq 0$ and ${\cal D} = 0$.  Eqns.(\ref{eq:2.11})
with  ${\cal D} = 0$  is seen, after some analysis, to be nothing more than  the DSE for the
constituent quark propagator  in the rainbow approximation (see  (\ref{eq:7}) and
(\ref{eq:8}) in sec.5).  The occurrence of the rainbow form of these equations is {\em not}
an approximation  within the GCM.  The non-rainbow diagrams, corresponding to
various more complicated gluon dressing of the quarks,  are generated by the additional
functional integrals in (\ref{eq:2.8}).  {\em Ad hoc} alterations to these DSE constituent
quark equations will lead to double counting of certain classes of diagrams.  The generation
of a minimum with ${\cal B}\neq 0$ is called the formation of a condensate, here a
$\overline{q}q$ condensate.  That ${\cal D} = 0$ means that in the GCM no diquark or
anti-diquark type condensates are formed. 

 Next in  going from (\ref{eq:2.7})  to (\ref{eq:2.8}) we must
consider the fluctuations or curvatures of the action for the bilocal fields. One finds
that  the curvature
$\delta^2 S_{bl}/\delta {\cal B}\delta {\cal B}$ when inverted gives the  meson
propagators, but with only ladder gluon exchanges.  Again non-ladder diagrams are
generated by the functional hadronic integrations in  (\ref{eq:2.8}).  Similarly inversion
of the curvatures in the diquark sector $\delta^2 S_{bl}/\delta {\cal D}\delta {\cal D^*}$
leads to diquark propagators, but with only ladder gluon exchanges between the constituent
quarks. 

We note that the generalised
bosonisation with meson and diquark fields leads to some additional complications that we
shall consider elsewhere, but which do not impinge on the basic  point being made here.
This meson-baryon hadronisation is based upon a generalised  Fierz transformation
\cite{RTC} that induces the appropriate  colour singlet anti-quark - quark correlations, and
colour anti-triplet quark-quark correlations that  are in the correct colour state for
quark-quark correlations within a colour singlet baryon.  

An earlier bosonisation
\cite{CR85}  used a Fierz transformation that lead to only the meson sector of the GCM. In
this bosonisation the meson effective action involves constituent states that are generated
by  naturally arising and exclusively rainbow or ladder diagrams. Then all of the other
diagrams contributing to the observable states are generated by the functional
integrations in (\ref{eq:2.8}).   Hence we see that in the exponent in (\ref{eq:2.8}) 
there arise  particular propagators for quarks, mesons, diquarks and even baryons. 
These propagators and their associated fields will be defined to be the {\em
constituent} states.    They could also be described as  core states.    The
observables are generated by the hadronic functional integrals in (\ref{eq:2.8}), and
correspond to dressing each constituent or core state with other such states.  Hence the
hadronic effective action in (\ref{eq:2.9}) contains a variety of parameters that refer to
the constituent states. 

 Nevertheless one often compares these parameters with the parameter
values for the fully dressed constituent states, that is the observable hadronic states.   
This appears to be valid because in general the dressing produces only a small shift in the
parameter values.  However one known exception is the nucleon where  pion dressing of the
constituent  nucleon state  reduces its mass by  some $200-300$MeV.  This mass shift
emerges from consideration of the functional integral 

\begin{equation} {\cal G}_{N}(X,Y)=\int{\cal D}\pi..
  \overline{N}(X)N(Y)exp(-S_{had}[\pi,...,\overline{N},N,..]), 
\label{eq:2.12}\end{equation}
where mainly the pions, but as well  other mesons are used to dress the nucleon.  Of
course one usually casts this into the form of a non-linear integral equation for the
meson dressed nucleon correlation function. 

\vspace{5mm}
\noindent {\bf 4. Chiral Limit  Constituent Pions} 

 When the  quark current masses ${\cal M} \rightarrow 0 \mbox{\ \ \ \ } 
S_{QCD}[\overline{q},q,A^a_{\mu}]$ has  an additional global 
\newline $U_L(N_F)\otimes U_R(N_F)$ chiral
symmetry:  writing 
$\overline{q}\gamma_{\mu}q=\overline{q}_R\gamma_{\mu}q_R+
\overline{q}_L\gamma_{\mu}q_L$  where $q_{R,L}=P_{R,L}q$ and
$\overline{q}_{R,L}=\overline{q}P_{L,R}$ we see that  these two   parts are separately
invariant under 
$q_R\rightarrow U_Rq_R,
\overline{q}_R\rightarrow\overline{q}_RU^{\dagger}_R$ and $q_L\rightarrow U_Lq_L,
\overline{q}_L\rightarrow\overline{q}_LU^{\dagger}_L$.  Its consequences may be explicitly
traced through the GCM hadronisation. First the ELEs $\delta S_{bl}/\delta {\cal B}$=0 have
degenerate solutions. In terms of the constituent quark propagator  this degeneracy
manifests itself  in the form
 \begin{equation}G(q;V)=[iA(q)q.\gamma+VB(q)]^{-1}=\zeta^\dagger G(q;{\bf
1})\zeta^\dagger\label{eq:2.13}\end{equation} where 
 \begin{equation}\zeta=\surd V, \mbox{\ \  }
V=exp(i\surd2\gamma_5\pi^aF^a)\label{eq:2.14}\end{equation}
in which the $\{\pi^a \}$   are arbitrary real constants.
The degeneracy of
the minimum implies that some   fluctuations in $\delta^2 S_{bl}/\delta {\cal B}\delta {\cal B}$
have zero mass; these are the  NG BSE states, and this indicates the realisation of the
Goldstone  theorem.

 In the hadronisation, in going from (\ref{eq:2.7}) to
(\ref{eq:2.8}),  new variables are forced upon us to describe the degenerate minima (vacuum
manifold). This is accomplished by a coordinatisation of the angle variables $\{\pi^a \}$:  
 \begin{equation}U(x)=exp(i\surd 2\pi^a(x)F^a)\end{equation} 
\begin{equation}V(x)=P_LU(x)^{\dagger}+P_RU(x) =exp(i\surd
2\gamma_5\pi^a(x)F^a)\label{eq:2.15}\end{equation}

  The NG part of the hadronisation then gives rise to the constituent pion effective action
\[
 \int d^4x\left( \frac{f_{\pi}^2}{4}tr(\partial_{\mu}U\partial_{\mu}U^
{\dagger})+\kappa_1tr(\partial^2U\partial^2U^{\dagger})+
\frac{\rho}{2}tr([{\bf 1} -\frac{U+U^{\dagger}}{2}]{\cal M})+\right.\]   
\begin{equation}\mbox{\ \ \ \ \ \ \ \ \ \ \ \ \ \ \ \ \ \ \ \ \ \ \ \ \ \ \ \ \ \ }
\left. +\kappa_2tr([\partial_{\mu}U\partial_{\mu}U^{\dagger}]^2)
+\kappa_3tr(\partial_{\mu}U\partial_{\nu}U^{\dagger}\partial_{\mu}U \partial_{\nu}U^
{\dagger})+....\right) \label{eq:2.16}\end{equation} 
This is the ChPT  effective action
\cite{CPT}, but with the added insight that all coefficients are given by explicit and
convergent integrals in terms of $A$ and
$B$, which are in turn  determined by $D_{\mu\nu}$.  The higher order terms contribute to
$\pi\pi$ scattering.  The dependence of the ChPT coefficients upon  $D_{\mu\nu}$ has been
studied  in \cite{Pi94,CG95a,Frank95} in which the GCM constituent pion expressions for
the various parameters were used.  However in view of the apparent generic role of the
GMOR relation one should keep in mind the possibility that the functional form of
the dependences of the parameters $\kappa_1,..$ on the quark correlation functions $A$ and
$B$ might also be generic. At present the final functional integral dressing  to obtain the
pion observables has not been  carried out. This amounts to the assumption  that the
constituent pion forms are sufficiently accurate.  The hadronisation procedure also gives a
full account of NG-meson - nucleon coupling.

The GCM is in turn  easily related to a number of the  more phenomenological models of QCD,
as indicated in  fig.1.  They include the Nambu-Jona-Lasinio Model (NJL) \cite{Reinhardt90},
ChPT \cite{CPT},  MIT and Cloudy Bag Model (CBM) \cite{Thomas}, Soliton Models \cite{Frank},
Quantum Hadrodynamics (QHD) \cite{QHD} and  Quantum Meson Coupling model (QMC) \cite{G88}.
We also indicate that the pure gluon correlation function in (\ref{eq:2.5}) may be
obtained from lattice computations and used in the GCM.  The relationships indicated in
fig.1 are discussed in \cite{CG96a}.
\vspace{5mm}
\begin{center}
\begin{picture}(10,150)(+200,80)
\thicklines
\put(5,185){\large \bf QCD}
\put(25,165){\vector(0,-1){30}}
\put(55,190){\vector(3,0){30}}
\put(95,185){\large \bf GCM}
\put(155,190){\vector(3,4){30}}
\put(200,225){\large \bf NJL}
\put(155,190){\vector(3,0){30}}
\put(200,185){\large \bf ChPT}
\put(155,190){\vector(3,-4){30}}
\put(200,155){\large \bf MIT, Cloudy Bag,  }
\put(200,140){\large \bf Soliton Models, QHD, QMC....}
\put(29,115){\large \bf Lattice Gluons}
\put(150,120){\vector(1,0){40}}
\put(120,135){\vector(0,1){30}}
\put(200,115){\large \bf Lattice Hadrons}
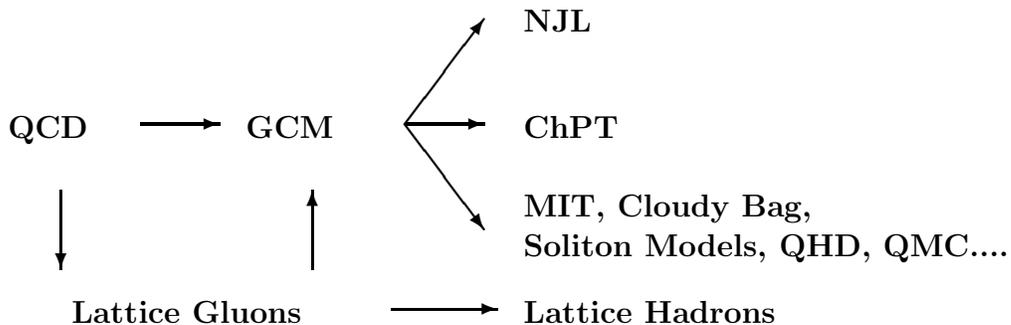
\put(10,90){  Fig.1.  Relationship of the GCM to QCD and other models}  
\end{picture}
\end{center}

\vspace{5mm}
\noindent {\bf 5. Action Minimum and Pionic Fluctuations}

The GCM involves the solution  of various integral equations for the  constituent
correlation functions. As we saw in sec.3,  the first equation involves the determination
of the minimum of the bilocal effective action, and this reduces to solving    the
DSE for the constituent quark propagator  necessarily in 
the rainbow form (the so-called vacuum equation of the GCM \cite{RTC,CR85})

\begin{equation}
B(p^2;m)=\frac{16}{3}\int\frac{d^4q}{(2\pi)^4}D(p-q).\frac{B(q^2;m)+m(q^2)}
{q^2A(q^2;m)^2+(B(q^2;m)+m(q^2))^2},
\label{eq:7}\end{equation}

\begin{equation}
[A(p^2;m)-1]p^2=\frac{8}{3}\int\frac{d^4q}{(2\pi)^4}q.pD(p-q).\frac{A(q^2;m)}
{q^2A(q^2;m)^2+(B(q^2;m)+m(q^2))^2},
\label{eq:8}\end{equation}
For simplicity we use a 
 Feynman-like gauge in which $D_{\mu\nu}(p)=\delta_{\mu\nu}D(p)$ (the quark-gluon coupling is
incorporated into $D$). The formal results of the analysis here are not gauge dependent.
Even in numerical studies  the Landau gauge can also be used; see \cite{CG95a}. We have
also included, for generality,  a running current mass for the quarks.

Using Fourier transforms (\ref{eq:7}) may be written in the form, here for $m=0$, 
\begin{equation}
D(x)=\frac{3}{16}\frac{B(x)}{\sigma_s(x)},
\label{eq:9}\end{equation}
which implies that knowledge of the quark propagator determines the effective GCM gluon
propagator. Multiplying (\ref{eq:9}) by $B(x)/D(x)$, and using  Parseval's identity for  the RHS,
we obtain the identity 
\begin{equation}
\int d^4x \frac{B(x)^2}{D(x)}=\frac{16}{3}\int\frac{d^4q}{(2\pi)^4}B(q)\sigma_s(q).
\label{eq:10}\end{equation}

The second basic equation is the ladder form BSE for the constituent pion mass-shell state,
which arises from the mesonic fluctuations about the minimum  determined by (\ref{eq:7}) and
(\ref{eq:8}). Again this ladder form cannot be generalised without causing double counting of
some classes of diagrams at a later stage, and without also damaging the intricate interplay
between  (\ref{eq:7}), (\ref{eq:8}) and the BSE
 \begin{equation}
\Gamma^f(p,P)=\frac{8}{3}\int\frac{d^4q}{(2\pi)^4}D(p-q)tr_{SF}(G_+T^gG_-T^f)\Gamma^g(q,P)
\label{eq:11}\end{equation}
where $G_{\pm}=G(q\pm\frac{P}{2})$. This BSE is for isovector NG bosons, and only the
dominant
$\Gamma=\Gamma^f T^f i\gamma_5$ amplitude is retained (see \cite{CRP87} for discussion); the spin
trace arises from projecting onto this dominant amplitude. Here $\{T^b,b=1,..,N_F^2-1\}$ are the
generators of $SU(N_F)$, with $tr(T^fT^g)=\frac{1}{2}\delta_{fg}$.  

The BSE (\ref{eq:11}) is an implicit
equation for the mass shell $P^2=-M^2$.  It  has solutions {\em only} in
the time-like region $P^2 \leq 0$. Fundamentally this is ensured by (\ref{eq:7}) and (\ref{eq:8})
being the specification of an absolute minimum of an effective action after a bosonisation
\cite {RTC}.  Nevertheless the loop momentum  is kept in the
space-like region $q^2 \geq 0$; this mixed metric device ensures that the
quark and gluon propagators remain close to the real space-like region where they have been most
thoroughly studied. Very little is known about these propagators in the time-like region $q^2 <
0$.  

The non-perturbative quark-gluon dynamics is expressed here in (\ref{eq:7}) and (\ref{eq:8}). Even
when  $m=0$  \hspace{2mm} (\ref{eq:7}) can have non-perturbative solutions with $B\neq 0$.
This is the dynamical breaking of chiral symmetry.

 When $m=0$  \hspace{2mm} (\ref{eq:11}) has a solution for $P^2=0$; the Goldstone theorem
effect. For the zero linear momentum  state  $\{P_0=0,\vec{P}=\vec{0}\}$  it is easily seen
that (\ref{eq:11}) reduces to (\ref{eq:7}) with $\Gamma^f(q,0)=B(q^2)$.  When $\vec{P}\neq
\vec{0} \mbox{\ \ } \mbox{\ then \ }  \Gamma^f(q,P)\neq B(q)$, and (\ref{eq:11})  must be solved 
for $\Gamma^f(q,P)$.

\vspace{5mm}
\noindent {\bf 6. Constituent  Pion  Mass}

We shall now determine an accurate expression for the mass of the constituent pion when
$m(s)$ is small but non-zero. This amounts to finding an analytic solution to the BSE
(\ref{eq:11}), when the constituent quark propagators are determined by (\ref{eq:7}) and
(\ref{eq:8}). The result will be accurate to order
$m$.

 For small $m \neq 0$
we can introduce the Taylor expansions in $m(s)$
\begin{equation}
B(s;m)+m(s) = B(s)+ m(s).\epsilon_s(s)+O(m^2),
\label{eq:12}\end{equation}
\begin{equation}
A(s;m) = A(s)+m(s).\epsilon_v(s)+O(m^2).
\label{eq:13}\end{equation}
 For large space-like $s$ we find  that $\epsilon_s
\rightarrow 1$, but for small $s$ we find that $\epsilon_s(s)$ can be significantly larger than
1. This is an infrared region dynamical enhancement  of the quark current mass by  gluon
dressing, and indicates the strong response of the chiral limit constituent quark
propagator to the turning on of the current mass.   A plot of
$\epsilon_s(s)$ is shown in
\cite{CG95b}.  Higashijima
\cite{Higashijima} and Elias \cite{Elias} have also reported similar enhancements of the
current quark masses in the infrared region.

 Even in the chiral limit  the constituent quark running mass 
$M(s)=B(s)/A(s)$  is essential for understanding any non-perturbative QCD quark effects. The
integrand of  a BSE contains the gluon correlation function,  constituent quark
correlation functions  and the form factor for the state (see for example (\ref{eq:11})).
This integrand shows strong peaking at typically  
$s \approx 0.3 GeV^2$. At this value we find \cite{CG95a} that $M(s) \approx 270MeV$.  This
is a property of the constituent hadrons.  It does not include any effects from the dressing
of these hadrons via  (\ref{eq:2.8}). This mass is called the constituent quark mass. 
Because of the  infrared region enhancement of the quark current mass we find that this
constituent mass rises quickly with quark current mass; see \cite{CG95a}.

Because the pion mass $m_{\pi}$ is small when $m$ is small, we can perform an expansion of
the
$P_{\mu}$ dependence in the  kernel of (\ref{eq:11}). Since the analysis is Lorentz covariant we
can, without loss of validity, choose to work in the rest frame with $P=(im_{\pi},\vec{0})$
giving, for  equal mass quarks for simplicity $$
\Gamma(p)=\frac{2}{9}m_{\pi}^2\int\frac{d^4q}{(2\pi)^4}D(p-q)I(s)\Gamma(q)+ 
\hspace{80mm} $$
\begin{equation}
\mbox{\ \ \ \ \ \ \ \ \ }
+\frac{16}{3}\int\frac{d^4q}{(2\pi)^4}D(p-q)\frac{1}{s(A(s)+m(s).\epsilon_v(s))^2+(B(s)+
m(s).\epsilon_s(s))^2}\Gamma(q)+....,
\label{eq:14}\end{equation}
where
\begin{equation}
I(s)=6\left(\sigma_v^2-2(\sigma_s\sigma_s'+s\sigma_v\sigma_v')-s(\sigma_s\sigma_s''-(\sigma_s')^2)
-s^2(\sigma_v\sigma_v''-(\sigma_v')^2)\right).
\label{eq:15}\end{equation}

By using Fourier transforms the  integral equation (\ref{eq:14}), now with explicit dependence on
$m_{\pi}$,  can be expressed in the form of a variational mass functional,
\begin{eqnarray*}
m_{\pi}[\Gamma]^2=-\frac{24}{f_{\pi}[\Gamma]^2}\int\frac{d^4q}{(2\pi)^4}\frac{\Gamma(q)^2}
{s(A(s)+m(s).\epsilon_v(s))^2+(B(s)+
m(s).\epsilon_s(s))^2}+\end{eqnarray*}
\begin{equation}\mbox{\ \ \ \ \ \ \ \ \ \ \ \ \ \ \ \ \ \ \ \ \ \ \ \ \ \ \ \ \ \ \ \ \ \ }+
\frac{9}{2f_{\pi}[\Gamma]^2}\int d^4x\frac{\Gamma(x)^2}{D(x)}
\label{eq:16}\end{equation}
in which
\begin{equation}
f_{\pi}[\Gamma]^2 =
\int\frac{d^4q}{(2\pi)^4}I(s)\Gamma(q)^2.
\label{eq:17}\end{equation}
The functional derivative $\delta m_{\pi}[\Gamma]^2/\delta\Gamma(q)=0$ reproduces
(\ref{eq:14}).  The mass functional  (\ref{eq:16}) and its  minimisation is equivalent to
the constituent pion BSE in the near chiral limit.  To find an estimate for the minimum we
need only note that the change in
$m_{\pi}^2$ from its chiral limit value of zero will be of 1st order in $m$, while the change
in the zero linear momentum frame $\Gamma(q)$ from its chiral limit  value $B(q^2)$ will be
of 2nd order in $m$.

Hence to obtain $m_{\pi}^2$ to lowest order in $m$, we may replace $\Gamma(q)$ by $B(q^2)$
in (\ref{eq:16}), and we have that the constituent pion mass is given by
\[
m_{\pi}^2=\frac{24}{f_{\pi}[B]^2}\int\frac{d^4q}{(2\pi)^4}m(s)\frac{\epsilon_s(s)B(s)+
s\epsilon_v(s)A(s)}{sA(s)^2+B(s)^2}
\frac{B(s)^2}{sA(s)^2+B(s)^2}
\mbox{\ \ \ \ \ \ \ \ \ \ \ \ \ \ \ \ \ \ \ \ \ \ \ \ \ \ \ \ \ \ \ } \]
\begin{equation}
-\frac{24}{f_{\pi}[B]^2}\int\frac{d^4q}{(2\pi)^4}
\frac{B(s)^2}{sA(s)^2+B(s)^2}+\frac{9}{2f_{\pi}[B]^2}\int
d^4x\frac{B(x)^2}{D(x)}+O(m^2)
\label{eq:18}\end{equation}
However the pion mass has been shown to be zero  in the chiral limit. This is confirmed
as the  two $O(m^0)$ terms in (\ref{eq:18})  cancel because of the identity (\ref{eq:10}). 
Note that it might appear that $f_{\pi}$ would contribute an extra $m$ dependence from its
kernel in (\ref{eq:15}).  However because the numerator in (\ref{eq:16}) is already of order
$m$, this extra contribution must be of higher order in $m$.  

Hence we  finally arrive at the  analytic expression, to $O(m)$, for the constituent NG boson
$(\mbox{mass})^2$ from the solution of the BSE in (\ref{eq:11})   

\begin{equation}
m_{\pi}^2=\frac{24}{f_{\pi}[B]^2}\int\frac{d^4q}{(2\pi)^4}m(s)\frac{\epsilon_s(s)B(s)+
s\epsilon_v(s)A(s)}{sA(s)^2+B(s)^2}
\frac{B(s)^2}{sA(s)^2+B(s)^2}+O(m^2).
\label{eq:19}\end{equation}
Eqn.(\ref{eq:20a})  or (\ref{eq:19}) is the new form of the NG mass formula derived in
\cite{CG95b}.  It would appear that expression (\ref{eq:20a}) 
is manifestly different to the conventional GMOR form in (\ref{eq:1}) and (\ref{eq:2}). 
However in the next section we generalise an  identity found by Langfeld and Kettner
\cite{Lang} which shows these forms to be equivalent.

\vspace{5mm}
\noindent {\bf 7. Relating the Mass Formulae}

Inserting (\ref{eq:12}) and (\ref{eq:13}) into (\ref{eq:7}), and expanding in powers of
$m(s)$,  we obtain up to terms linear in
$m$, and after using  (\ref{eq:7}) with $m=0$ to eliminate the $O(m^0)$ terms,

$$
m(p^2)\epsilon_s(p^2)=m(p^2)+\frac{16}{3}\int\frac{d^4q}{(2\pi)^4}D(p-q)
\frac{m(q^2)\epsilon_s(q^2)}{q^2A(q^2)^2+B(q^2)^2}\mbox{\ \ \ \ \ \ \ \ }$$
$$-\frac{16}{3}\int\frac{d^4q}{(2\pi)^4}D(p-q)\frac{B(q^2)^22m(q^2)
\epsilon_s(q^2)}{(q^2A(q^2)^2+B(q^2)^2)^2}$$
\begin{equation}\mbox{\ \ \ \ \ \ \ \ \ \ \ \ \ \ \ \ \ \ \ \ \ \ \ \ \ \ \ \ }
-\frac{16}{3}\int\frac{d^4q}{(2\pi)^4}D(p-q)
\frac{B(q^2)A(q^2)2m(q^2)q^2\epsilon_v(q^2)}{(q^2A(q^2)^2+B(q^2)^2)^2}.
\label{eq:GenLK}\end{equation}

We now multiply (\ref{eq:GenLK}) throughout by  $B(p^2)/(p^2A(p^2)^2+B(p^2)^2)$,
and integrate wrt $p$.  Using again the chiral limit of (\ref{eq:7}) there is some
cancellation of terms, and we are left with a generalised Langfeld-Kettner identity

$$2\int d^4p
\frac{B(p^2)^2}{p^2A(p^2)^2+B(p^2)^2}\left(\frac{B(p^2)m(p^2)\epsilon_s(p^2)}
{p^2A(p^2)^2+B(p^2)^2}+\frac{p^2A(p^2)m(p^2)\epsilon_v(p^2)}{p^2A(p^2)^2+B(p^2)^2}\right)=
\mbox{\ \ \ \ \ \ \ \ }$$
\begin{equation}\mbox{\ \ \ \ \ \ \ \ \ \ \ \ \ \ \ \ \ \ \  }
\mbox{\ \ \ \ \ \ \ \ \ \ \ \ \ \ \ \ \ \ \ \ \ \ \ \ \ \ \ \ }
\int d^4p\frac{m(p^2)B(p^2)}{p^2A(p^2)^2+B(p^2)^2}
\label{eq:ident}
\end{equation}
Remarkably, noting (\ref{eq:3a}) and (\ref{eq:3b}), we see that using this identity
in  (\ref{eq:20a}) or (\ref{eq:19}) finally completes the derivation of the GMOR expression
for the mass of the constituent pion. 

 We thus see that despite its apparently simple form 
the GMOR expression  actually depends  on two identities that follow from the {\em non-linear}
constituent quark DSE, as well as on the  subtle interplay between this  constituent quark
equation and the BSE  for the constituent pion.  These in turn arise from the careful
self-consistency rendered by the functional integral prescription which ensures that the 
fluctuation spectrum for the bilocal action is precisely related to the Euler-Lagrange
equations. {\em Ad hoc} alterations   will invalidate this connection and  therefore the  
derivation of the GMOR expression.

\vspace{5mm}

\noindent {\bf 8. Conclusion}  

We have indicated the careful considerations that must be given to modelling QCD via the
GCM and the manner in which this leads to hadronic effective actions.  We have defined
constituent quarks, meson, diquarks and baryons as those states that appear in the
effective action, i.e. in the exponent, as in (\ref{eq:2.8}).  These constituent states
are then further dressed by the functional integrations in (\ref{eq:2.8}). Remarkably this
GCM structuring of the quantum field theoretic  analysis implies, at least in the simplest
version of the GCM, that the constituent states are described by sums of rainbow or
ladder diagrams, and that the functional integrations then build up all the remaining
diagrams, amounting to the vast array of crossed diagrams and vertex dressings etc.  Because
in most cases  these extra dressings do not cause large changes in the values of
the constituent  masses, coupling constants,..  the effect of  the inclusion
of these extra diagrams is not manifestly large.  Of course this is not surprising because
the GCM hadronisation allows us to assess the significance of a constituent state through its
mass; low mass states should be more important than very massive states.  This implies that
the pion dressing is the largest such effect. However inclusion of this dressing for the
constituent nucleon is known to be significant, and is a result of the inclusion of various
non-ladder diagrams in the observable  nucleon. 

We have also carefully indicated that it is the mass of the {\em constituent} pion that is
analysed here and, by using various identities that follow from the non-linear equation for
the constituent quark,  one can show that the mass of this constituent pion is indeed given
by the GMOR formula, with the scalar part of the constituent quark correlation function
appearing. This does not preclude the fact that presumably the observable pion also has its
mass  obeying a GMOR formula, but one in which the full quark  scalar correlation function
appears.   That is, the GMOR relation is  generic.


\newpage



 
\begin{thebibliography}{99}

\bibitem{GMOR} M. Gell-Mann, R. Oakes, and B. Renner, Phys. Rev. {\bf 175}(1968)2195.

\bibitem{CG95b} R. T. Cahill and S. M. Gunner, Mod. Phys. Lett. A {\bf 10}(1995)3051.

\bibitem{FR96}  M. R. Frank and C. D. Roberts, Phys. Rev. C {\bf 53}(1996)390.

\bibitem{Lang} K. Langfeld and  C. Kettner, {\em The quark condensate in the GMOR relation},
hep-ph/9601370.


\bibitem{OPE} M. A. Shifman, A. I. Vainshtein, V. I. Zakharov, Nucl. Phys. {\bf
B147}(1979)385, V. A. Novikov, M. A. Shifman, A. I. Vainshtein and V. I. Zakharov, Nucl.
Phys. {\bf B174}(1980)378, Nucl. Phys. {\bf B249}(1985)445.

\bibitem{Rein85} L. J. Reinders, H. Rubinstein and S. Yazaki, Phys. Rep. {\bf 127}(1985)1.

\bibitem{Nar89} S. Narison, {\em QCD Spectral Sum Rules}, World Scientific Lecture Notes in
Physics Vol. 26, Singapore 1989.

\bibitem{Bijnens} J. Bijnens, J. Prades and E. de Rafael, Phys. Lett. B {\bf 348}(1995)226.


\bibitem{RTC} R. T. Cahill,  Nucl. Phys. A {\bf 543}(1992)63c;
                             Aust. J. Phys. {\bf 42}(1989)171.

\bibitem{Pennington} N. Brown and M. R. Pennington, Phys. Rev. D {\bf 39}(1989)2723;
             K. B\"{u}ttner and M. R. Pennington, Phys. Rev. D {\bf 52}(1995)5220.

\bibitem{Marenzoni93} P. Marenzoni, G. Martinelli, N. Stella and M. Testa, Phys. Lett. B
                     {\bf 318} (1993)511. 

\bibitem{CG95a}  R. T. Cahill and S. M. Gunner, Phys. Lett. B {\bf 359}(1995)281.

\bibitem{CR85} R. T. Cahill and C. D. Roberts, Phys. Rev. D {\bf 32}(1985)2419.

\bibitem{CPT} J. Gasser and H. Leutwyler, Ann. Phys.(NY) {\bf 158}(1984)142; Nucl. Phys. B
{\bf 250}(1985)465.

\bibitem{Pi94}  C. D. Roberts, R. T. Cahill, M. E. Sevior and N. Iannella,
                Phys. Rev. D {\bf 49}(1994)125.

\bibitem{Frank95}  M. R. Frank and  T. Meissner, {\em Low-energy QCD: Chiral coefficients and
the quark-quark interaction}, nucl-th/9511016.


\bibitem{Reinhardt90} H. Reinhardt, Phys. Lett. B {\bf 244}(1990)316.

\bibitem{Thomas}  A. W. Thomas, S. Theberge and G. A. Miller, Phys. Rev. D {\bf
24}(1981)216.

\bibitem{Frank} M. R. Frank and P. C. Tandy, Phys. Rev. C {\bf 46}(1992)338.

\bibitem{QHD}  B. D. Serot, Rep. Prog. Phys. {\bf 55}(1992)1855.

\bibitem{G88}  P.A.M. Guichon, Phys. Lett. B {\bf 200}(1988)235.

\bibitem{CG96a} R. T. Cahill and S. M. Gunner, {\em The global colour model of QCD and its 
      relationship to the NJL model, chiral perturbation theory and other models},
      hep-ph/9601319.

\bibitem{CRP87} R. T. Cahill, C. D. Roberts and J. Praschifka, Phys. Rev. D {\bf
36}(1987)2804.

\bibitem{Higashijima} K. Higashijima, Phys. Rev. {\bf 29}(1984)1228.

\bibitem{Elias} V. Elias, Can. J. Phys. {\bf 71}(1993)347. 



 \end{thebibliography}
\end{document}